\documentclass[journal]{IEEEtran}
\pdfoutput=1
\ifCLASSINFOpdf
   \usepackage[pdftex]{graphicx}
\else
  \usepackage[dvips]{graphicx}
\fi

\usepackage{subfigure}
\usepackage{graphicx}
\usepackage{mathptmx}
\usepackage{latexsym}
\usepackage{amssymb}
\usepackage{amsmath}
\usepackage{epstopdf}
\usepackage{colortbl}
\usepackage{multirow}
\usepackage{makecell}
\usepackage{amsfonts}
\usepackage{url}
\allowdisplaybreaks

\begin{document}

\title{Secure-ISAC: \underline{Secure} V2X Communication: An \underline{I}ntegrated \underline{S}ensing \underline{a}nd \underline{C}ommunication Perspective}

\author{Kan Yu,~\IEEEmembership{Member, IEEE}, Zhiyong~Feng,~\IEEEmembership{Senior Member,~IEEE}, Dong Li,~\IEEEmembership{Senior Member, IEEE}, and Jiguo Yu,~\IEEEmembership{Fellow,~IEEE}
%, Xiangyun Zhou,~\IEEEmembership{Fellow,~IEEE}, and Ping Zhang,~\IEEEmembership{Fellow,~IEEE}

\thanks{This work is supported by the Macao Young Scholars Program with Grants AM2023015, the National Natural Science Foundation of China with Grants 62301076, 62272256 and 62202250, the Natural Science Foundation of Shandong Province with Grants ZR2021QF050 and ZR2021MF075, Major Program of Shandong Provincial Natural Science Foundation for the Fundamental Research under Grant ZR2022ZD03, the Pilot Project for Integrated Innovation of Science, Education and Industry of Qilu University of Technology (Shandong Academy of Sciences) under Grant 2022XD001, the Talent Cultivation Promotion Program of Computer Science and Technology in Qilu University of Technology (Shandong Academy of Sciences) under Grant 2023PY059, the Colleges and Universities 20 Terms Foundation of Jinan City with Grant 202228093, and the Science and Technology Development Fund, Macau SAR, under Grant 0029/2021/AGJ.}
\IEEEcompsocitemizethanks{\IEEEcompsocthanksitem
K. Yu is with the School of Computer Science and Engineering, Macau University of Science and Technology, Taipa, Macau, 999078, P. R. China; the Key Laboratory of Universal Wireless Communications, Ministry of Education, Beijing University of Posts and Telecommunications, Beijing, 100876, P.R. China. E-mail: kanyu1108@126.com. \protect\\
Z. Feng is with the Key Laboratory of Universal Wireless Communications, Ministry of Education, Beijing University of Posts and Telecommunications, Beijing, 100876, P.R. China. E-mail: fengzy@bupt.edu.cn.\protect\\
%D. Ma is the Key Laboratory of Universal Wireless Communications, Ministry of Education, Beijing University of Posts and Telecommunications, Beijing, 100876, P.R. China. E-mail: dingyouma@bupt.edu.cn.\protect\\
D. Li (the corresponding author) is with the School of Computer Science and Engineering, Macau University of Science and Technology, Taipa, Macau, 999078, P. R. China. E-mail: dli@must.edu.mo.\protect\\
J. Yu (Co-corresponding author) is with Big Data Institute, Qilu Univeristy of Technology, Jinan, Shandong, 252353, P.R. China. E-mail: jiguoyu@sina.com.\protect\\
%X. Zhou is with the School of Engineering, The Australian National University, Canberra, ACT 2601, Australia. E-mail: xiangyun.zhou@anu.edu.au.\protect\\
%P. Zhang is with the State Key Laboratory of Networking and Switching Technology, Beijing University of Posts and Telecommunications, Beijing 100876, P. R. China, E-mail: pzhang@bupt.edu.cn.\protect\\
%D. Yu is with School of Computer Science and Technology, Shandong University, Qingdao, 266510, P.R. China. E-mail: dxyu@sdu.edu.cn.\protect\\
%Y. Zou is with School of Computer Science and Technology, Shandong University, Qingdao, 266510, P.R. China. E-mail: yfzou@sdu.edu.cn.\protect\\
}
}

\markboth{IEEE Wireless Communications,~Vol.~ , No.~ , 2023}%
{Shell \MakeLowercase{\textit{et al.}}: Bare Demo of IEEEtran.cls for Computer Society Journals}
% The only time the second header will appear is for the odd numbered pages
% after the title page when using the twoside option.
%
% *** Note that you probably will NOT want to include the author's ***
% *** name in the headers of peer review papers.                   ***
% You can use \ifCLASSOPTIONpeerreview for conditional compilation here if
% you desire.

% The publisher's ID mark at the bottom of the page is less important with
% Computer Society journal papers as those publications place the marks
% outside of the main text columns and, therefore, unlike regular IEEE
% journals, the available text space is not reduced by their presence.
% If you want to put a publisher's ID mark on the page you can do it like
% this:
%\IEEEpubid{0000--0000/00\$00.00~\copyright~2014 IEEE}
% or like this to get the Computer Society new two part style.
%\IEEEpubid{\makebox[\columnwidth]{\hfill 0000--0000/00/\$00.00~\copyright~2014 IEEE}%
%\hspace{\columnsep}\makebox[\columnwidth]{Published by the IEEE Computer Society\hfill}}
% Remember, if you use this you must call \IEEEpubidadjcol in the second
% column for its text to clear the IEEEpubid mark (Computer Society jorunal
% papers don't need this extra clearance.)

% use for special paper notices
%\IEEEspecialpapernotice{(Invited Paper)}

% for Computer Society papers, we must declare the abstract and index terms
% PRIOR to the title within the \IEEEtitleabstractindextext IEEEtran
% command as these need to go into the title area created by \maketitle.
% As a general rule, do not put math, special symbols or citations
% in the abstract or keywords.
\IEEEtitleabstractindextext{%
\begin{abstract}
In Vehicle-to-Everything (V2X) systems, reliable and secure information exchange  plays a pivotal role in road safety and traffic management. Due to the open nature of the wireless medium and the constant or intermittent mobility of vehicles, the security of transmissions in V2X is more challenging compared to traditional wireless networks. Physical layer security (PLS) leverages the inherent randomness of wireless communication channels to ensure the confidentiality and security of information transmission. Current PLS schemes in integrated communications and sensing (ISAC) enabled V2X systems is to utilize communication interference to significantly impact the eavesdropping channel more than the legitimate channel. However, in an ISAC-enabled V2X system, it is crucial to prioritize and address the issue of interference coupling as it significantly impacts the confidentiality and security of information exchange. This goes beyond simply relying on the communication interference. Until now, no discussions or studies on integrating security with ISAC (Seucue-ISAC) in ISAC-enabled V2X systems, specifically regarding the exploitation of sensing interference or coupling interference. In this article, we provide a comprehensive review on PLS metrics and security threats encountered in V2X communication. And then, we discuss and analyze four popular PLS techniques and the main challenges associated with their implementation in ISAC-enabled V2X systems. Finally, we share our vision for PLS studies in ISAC-based V2X systems to promote Secure-ISAC.

\end{abstract}

% Note that keywords are not normally used for peerreview papers.
\begin{IEEEkeywords}
Vehicle-to-Everything; Physical layer security; Integrated security with ISAC; Sensing interference
\end{IEEEkeywords}}

% make the title area
\maketitle

% To allow for easy dual compilation without having to reenter the
% abstract/keywords data, the \IEEEtitleabstractindextext text will
% not be used in maketitle, but will appear (i.e., to be "transported")
% here as \IEEEdisplaynontitleabstractindextext when the compsoc
% or transmag modes are not selected <OR> if conference mode is selected
% - because all conference papers position the abstract like regular
% papers do.
\IEEEdisplaynontitleabstractindextext
% \IEEEdisplaynontitleabstractindextext has no effect when using
% compsoc or transmag under a non-conference mode.

% For peer review papers, you can put extra information on the cover
% page as needed:
% \ifCLASSOPTIONpeerreview
% \begin{center} \bfseries EDICS Category: 3-BBND \end{center}
% \fi
%
% For peerreview papers, this IEEEtran command inserts a page break and
% creates the second title. It will be ignored for other modes.
\IEEEpeerreviewmaketitle

\section{Introduction}\label{sec:intro}
Recently, vehicle-to-everything (V2X) communication has emerged as a transformative paradigm, connecting vehicles with every aspect of the environment. It has ushered in a new era for intelligent transportation systems (ITS), bolstered by the support of artificial intelligence \cite{SiegelTITS2018,AbbasTITS2019}. Achieving constant connectivity, stringent reliability, low-latency guarantees, and efficient resource allocation for information exchange in V2X necessitates highly efficient coordination of vehicular communications and sensing.
The integrated sensing and communications (ISAC) technique plays a pivotal role in enhancing road safety and traffic efficiency within ITS through the seamless integration of communications and sensing \cite{Zhang2023WC}. Specifically, leveraging the precise localization capabilities and narrow-pointing beamforming of vehicular/automotive radar operating in the millimeter-wave frequency range allows for the acquisition of vast sensing data. This, in turn, facilitates vehicular communications. Furthermore, with the help of narrow beamwidth and ample bandwidth of millimeter-wave technology, communication among vehicles not only caters to diverse information exchange requirements, such as vehicle-to-vehicle (V2V), vehicle-to-pedestrian (V2P), and vehicle-to-infrastructure (V2I), each with unique demands like ultra-low latency and extensive bandwidth for navigation map loading, but also enables ``sensing beyond the visible'' by rapidly sharing and fusing sensing data.
In summary, ensuring reliable and timely information interaction between sensing data and communication data is paramount.

Given the inherent openness of the wireless medium, ensuring the confidentiality and security of information becomes paramount \cite{Yu2021tvt,YuToN2023}. Traditionally, security concerns, operating under the assumption of limited computational power on the part of eavesdroppers, have been addressed by employing encryption methods in the network layer or higher layers of the protocol stack.
Nevertheless, applying this security paradigm to secure V2X communication proves to be inefficient for the following reasons: 1) The concept of perfect secrecy can be compromised when eavesdropping vehicles possess sufficient computational power, enabled by artificial intelligence and quantum computation; 2) Secret key management becomes more challenging due to vehicular mobility.
%\begin{enumerate}
% \item The concept of perfect secrecy can be compromised when eavesdropping vehicles possess sufficient computational power, enabled by artificial intelligence and quantum computation.
%  \item Apart from broadcasting wireless channel, secret key management becomes more challenging due to vehicular mobility.
%\end{enumerate}
Given this context, growing concerns have been raised regarding the security of traditional cryptography methods, necessitating the urgent development of more effective techniques to enhance V2X communication security. From a physical layer perspective, the presence of noise, fading, and unintended signal interference in the wireless channel is often considered as impairments. Differing from traditional encryption methods, physical layer security (PLS) offers lower complexity and reduced end-to-end latency. It leverages the characteristics of the physical layer to conceal transmitted messages, particularly when the quality of the legitimate channel surpasses that of the eavesdropping channel. In conjunction with traditional encryption schemes, PLS ensures comprehensive security for V2X communication, spanning from the physical layer to the application layer of the protocol stack. It's worth noting that in the context of an ISAC and millimeter-wave-based V2X system, there is an increasing overlap between the frequency bands used for communications and sensing. This overlap leads to complex coupling interference and a performance trade-off between the two, which can pose challenges.

Statistics indicate that when vehicles are navigating intersections, changing lanes, or following other vehicles, there is approximately a 75\% probability of encountering interference.
There are two primary factors to be considered when addressing interference in the context of communications and sensing. Firstly, unlike the broadcast nature of wireless communication interference, interference arising from sensing beamforming occurs when vehicular radars employ beamforming within each other's sensing range, resulting in overlapping pointing beams. In such cases, this interference can lead to a degradation in the psensing performance. Secondly, the fundamental principle behind current PLS schemes in an ISAC-enabled V2X is the utilization of wireless communication interference, including noise and fading, to significantly impact the eavesdropping channel as compared to the legitimate channel.
In fact, it is crucial to carefully consider coupling interference in ISAC-enabled V2X systems to ensure the confidentiality and security of information exchange. This goes beyond solely relying on communication interference. To date, there has been a lack of research on PLS in ISAC-enabled V2X systems that share the millimeter-wave band with vehicular communication and vehicular sensing radar, particularly with respect to exploiting sensing interference or coupling interference.

\begin{figure}[!ht]
\centering
\includegraphics[width=3.1in]{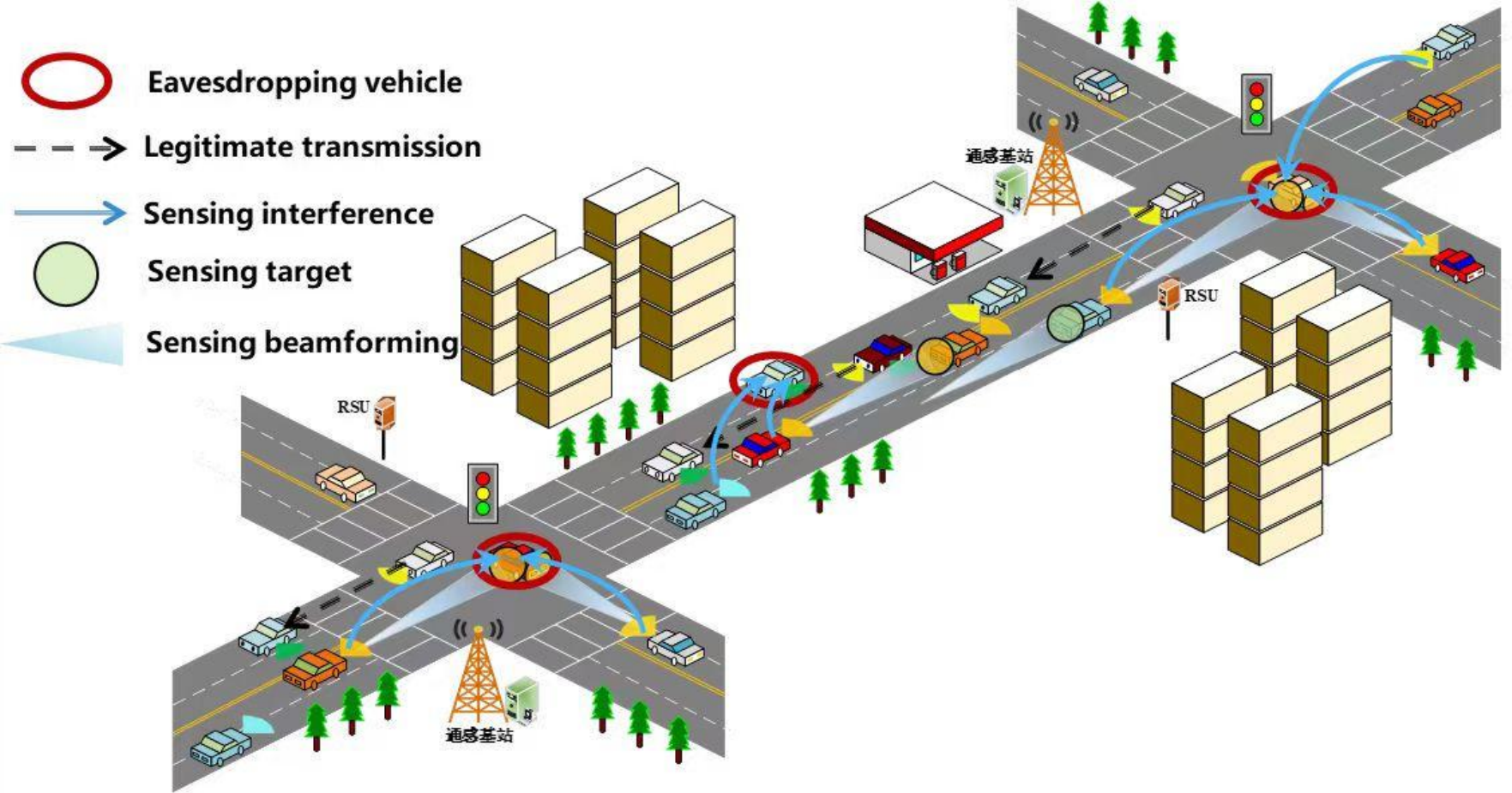}
\caption{\small The PLS enhancement through exploiting sensing beamformings in ISAC-enabled V2X}
\label{fig:scenario_all}
\end{figure}

As illustrated in Fig. \ref{fig:scenario_all}, when the malicious vehicle is closer to the legitimate transmitting vehicle than the receiver, the sensing beamforming can serve as interference, effectively thwarting eavesdropping vehicles instead of causing communication interference. In addition, the interference resulting from sensing beamforming has detrimental effects on the reliability of V2X communication. However, the precise relationship among communication reliability, sensing accuracy, and security of physical layer remains unclear. Hence, the task of depicting the interaction effects and harnessing coupling interference between communications and sensing for secure V2X communication presents a significant challenge. Furthermore, the integration of security with ISAC imposes greater advantages on the secrecy performance of ISAC-enabled V2X.

In this article, we focus on the PLS efficiency improvement of an IASC-enabled V2X system. First, we analyze the popular PLS metrics and security threats of V2X.
Then, the available PLS techniques adopted for secure V2X communication are summarized from the perspective of trade-off between reliability and security, and the major challenges to implement them in an ISAC-enabled V2X system are discussed. Furthermore, we discuss the open issues about PLS of IASC-based V2X systems and their potential implications in V2X communications, followed by the conclusions and future works.

\section{Overview of PLS Strategies in Wireless Communications}\label{sec:Overview of PLS Strategies}

\subsection{PLS Performance Metrics}
In this section, we delve into the examination of key security metrics used to evaluate the reliability and security of PLS-based systems, emphasizing their relevance within the context of information theory (i.e., infinite block-length). Moreover, ultra-reliable and low-latency communication (uRLLC) usually necessitates finite block-length to achieve low-latency requirement, under which the decoding error probability (DEP) cannot be underestimated when compared to information theory. As a result,  We also introduce a set of metrics tailored for finite block-length, which are crucial for gauging the security performance of PLS-based uRLLC.

From an information-theoretic perspective, Wyner demonstrated that perfect secrecy can be attained when the legitimate receiver possesses superior channel quality relative to all eavesdroppers. This insight gives rise to four primary security metrics used for evaluating the performance of PLS-based systems \cite{Wyner1975}, which include the connection outage probability, secrecy outage probability, and secrecy rate maximization, defined as follows:
\begin{itemize}
  \item \emph{Connection outage probability}, which assesses reliability, quantifies the probability that the rate of the legitimate channel falls below a specific threshold value.
  \item \emph{Secrecy outage probability}, which evaluates security, represents the probability that the rate of the eavesdropping channel exceeds a certain threshold value.
  \item \emph{Secrecy rate maximization} involves the optimization of the non-negative rate difference between the achievable rates of the legitimate receiver and the eavesdropper. This metric characterizes the maximum secrecy rate at which information can be transmitted securely and reliably. In addition, secrecy rate maximization may entail constraints related to the expected levels of reliability and security.
\end{itemize}

\textbf{Finite block-length:} Different from the perspective of information theory, when block-length is finite, DEP at the receiver is inevitable even if the SINR condition is satisfied. Moreover, the probability can be determined by $f\left(\frac{\log_2(1+\gamma) - R}{\sqrt{V/n}}\right)$, where $R$ denotes the rate at the legitimate receiver and eavesdropper, $V=(\log_2e)^2[1-1/(1+\gamma)^2]$ is the channel dispersion, and $f(x)=\int^{+\infty}_{x}e^{-t^2/2}\sqrt{2\pi}dt$ denotes the Q-function \cite{Sun2018}. Considering the non-negligible decoding error, there are four main security metrics for evaluating the performance in finite block-length PLS systems. They are the security gap, rate interval, BER-COP, and BER-SOP, secrecy rate maximization under finite block-length, which are defined as follows.

\begin{itemize}
  \item \emph{The security gap} is the ratio of $SNR_{b,\min}$ and $SNR_{e, \max}$, where $SNR_{b,\min}$ is the minimum received SNR at the legitimate receiver when the corresponding bit error rate (BER) caused by the finite block-length is less than the expected value, $SNR_{e, \max}$ is the maximum received SNR at the eavesdropper while the corresponding BER is greater than the expected value. In this sense, security gap describes the difference of legitimate and wiretap channels constrained by the BER.
  \item \emph{The rate interval} refers to the difference between the maximum rate of legitimate receiver and the minimum rate, under which the BER at legitimate receivers is less than the desired value, while the BER at eavesdroppers is over the desired value.
  \item \emph{The BER-COP} and \emph{BER-SOP}. Let $\epsilon^{\mathrm{max}}_{u,\mathrm{BER}}$ and $\epsilon^{\mathrm{min}}_{e,\mathrm{BER}}$ denote the acceptable maximum and minimum DEPs for the expected reliability and security. Let $\beta_{u,\mathrm{min}}$, the reliability threshold, be the lowest SINR at legitimate users satisfying the reliability constraint that $\epsilon_{u,\mathrm{BER}}$ is less than or equal to $\epsilon^{\mathrm{max}}_{u,\mathrm{BER}}$, and $\beta_{e,\mathrm{max}}$, the security threshold, be the highest SINR at any Eves, and corresponding $\epsilon_{e,\mathrm{BER}}$ is greater than or equal to $\epsilon^{\mathrm{min}}_{e,\mathrm{BER}}$. Accordingly, based on the COP and SOP with the infinite block-length, denoted by $ p_{co}$ and $p_{so}$, the BER-COP and BER-SOP can be represented as $\epsilon^{\mathrm{max}}_{u,\mathrm{BER}}(1 - p_{co}) + p_{co}$ and $ (1 - \epsilon^{\mathrm{min}}_{e,\mathrm{BER}}) p_{so}$, respectively.
  \item \emph{The secrecy rate maximization} is to maximize the non-negative rate difference between the achievable rates of the legitimate receiver and the eavesdropper, under the constraint of the expected maximum BER-COP and minimum BER-SOP for reliability and security, rather than the constraint of the expected level of reliability and security with the infinite block-length.
\end{itemize}

\subsection{Attack Threats in Wireless Communications}
\emph{Passive attack}: In V2X communication, passive attacks can occur when potential malicious vehicles are driving on the road in the same direction as the transmitting and receiving vehicles. In such cases, the channel quality of eavesdropping vehicles can be significantly superior to that of the receiving vehicle. This is because the beamforming is directed towards the malicious vehicles, while the line-of-sight (LoS) link between legitimate vehicles is further obstructed by these malicious vehicles.

\emph{Active attack}: an adversary vehicle transmits non-intended signals to interfere the receiving vehicles, especially during critical delay-sensitive maneuvers between legitimate parties in V2X communications. Moreover, accurate CSI of legitimate vehicles for effective beamforming usually is obtained through channel training with pilot signals. In this case, active malicious vehicles can impersonate legitimate vehicles by transmitting the same pilot signals, deceiving legitimate vehicles, and facilitating more effective eavesdropping.

\begin{figure*}
\centering
\subfigure[\small design of beamforming and trajectory based on UAV]{
\centering
\includegraphics[width=2.8in]{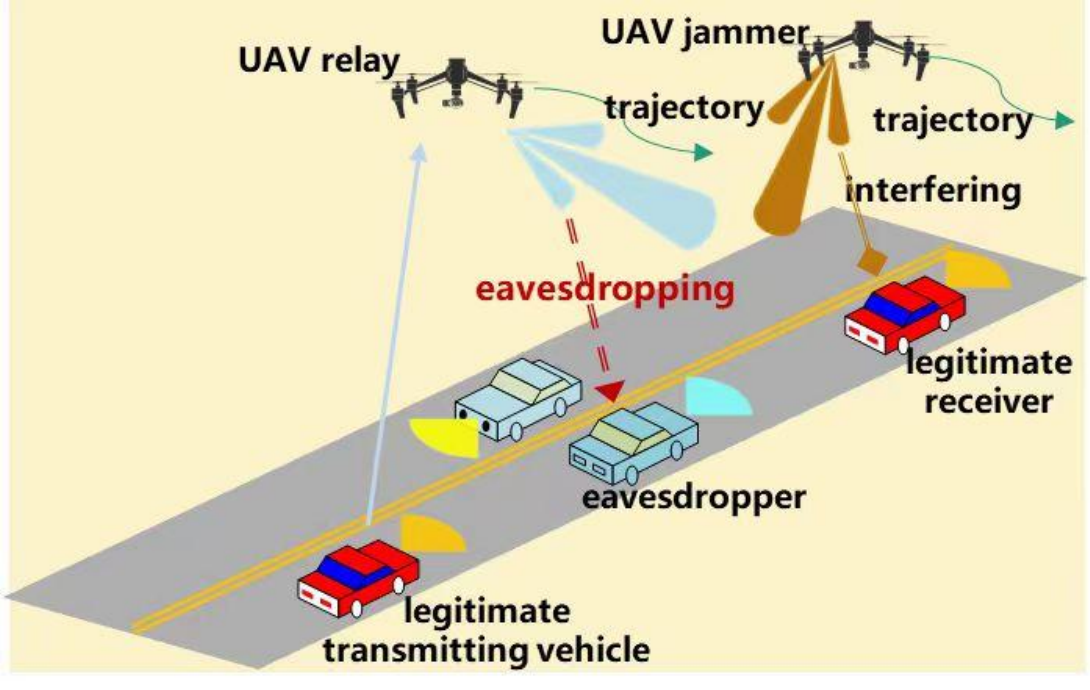}
\label{fig:scenario3}}
\subfigure[\small design of vehicular beamforming and trajectory]{
\centering
\includegraphics[width=2.8in]{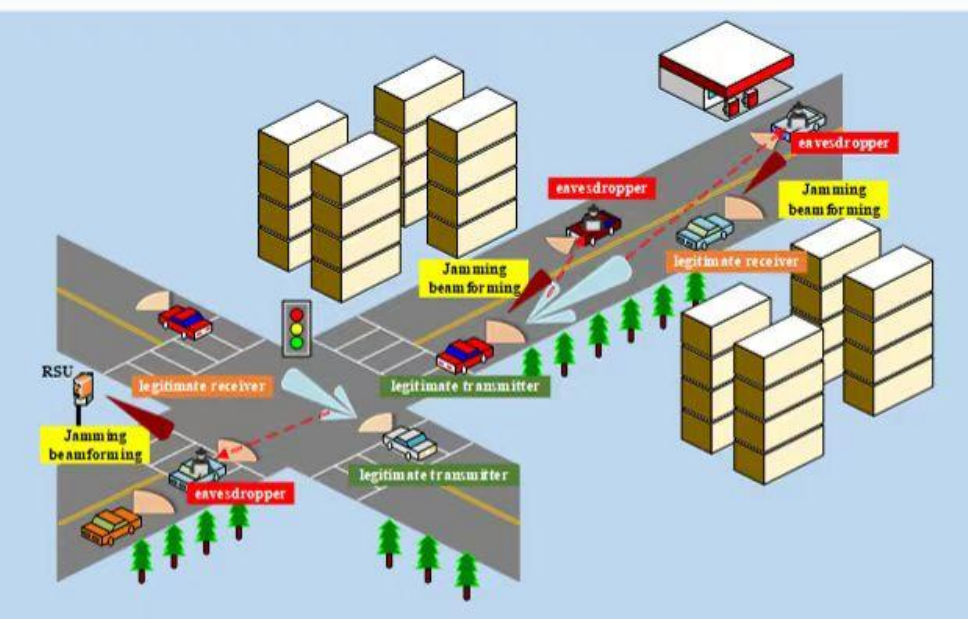}\label{fig:scenario4}}
\subfigure[\small design of beamforming and RIS-based phase shift]{
\centering
\includegraphics[width=2.8in]{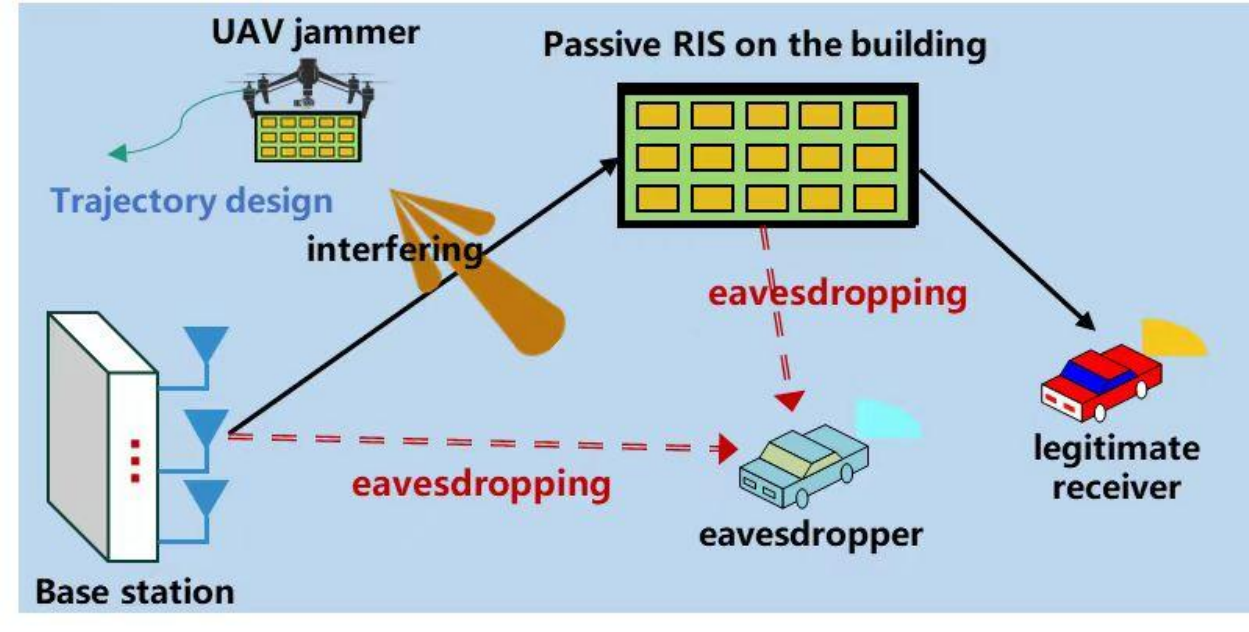}\label{fig:scenario1}}
\subfigure[\small design of RIS-based beamforming and UAV trajectory]{
\centering
\includegraphics[width=2.9in]{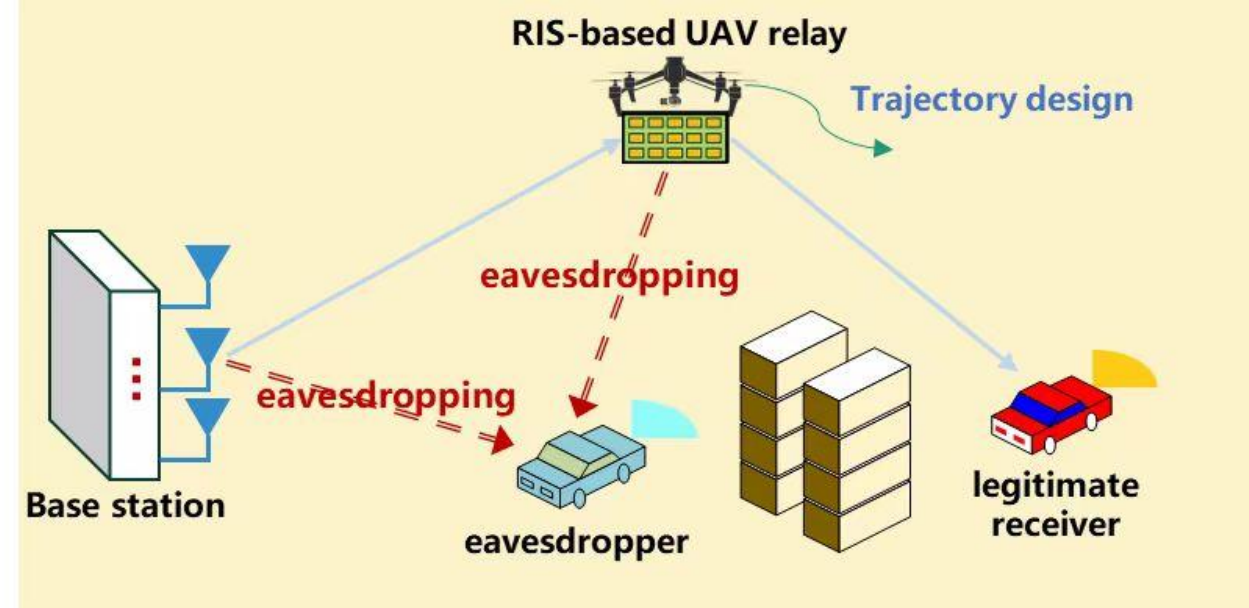}\label{fig:scenario2}}
\caption{\small Joint optimization design of UAV trajectory, vehicular trajectory, and RIS-based beamforming.}
\end{figure*}

\section{Security-Reliability Tradeoff}\label{sec:security reliability tradeoff}
As previously discussed, stringent ultra-reliable, low-latency, and higher security are crucial needs in V2X communication. The proper design of transmit power for legitimate vehicles is critical for finite block-length, as an inadequate setting may result in weak signals that are easily overwhelmed by non-intended signals. On the other hand, increasing transmit power can enhance the signal strength (and decrease DEP caused by block-length) but may also inadvertently facilitate eavesdropping. Therefore, striking the right balance between security and reliability is of paramount importance.
Relay systems, including UAV, RSU and other vehicles, have proven to be valuable for improving reliability, especially in scenarios with unfavorable channel conditions \cite{Rodrguez2015CM}.
Additionally, cooperative jamming schemes offer another avenue for enhancing security. These schemes aim to significantly suppress eavesdropping channels relative to legitimate channels, often by deliberately generating Artificial Noise (AN) to confound eavesdropping vehicles \cite{Yin2022TITS}.

To ensure a positive secrecy rate at all times, many current works assumed that the channel gain from the legitimate vehicles to eavesdropping vehicles was always less than that from the legitimate vehicles. While this assumption is reasonable in numerous cases, it doesn't always hold in V2X communication due to potential obstacles and Non-Line-of-Sight (NLoS) links caused by vehicle mobility. Therefore, we consider a more practical scenario in which channel gains are truly independent, and malicious vehicles have a better channel than the receiving vehicles. We then provide a comprehensive overview of PLS strategies employed for secure V2X communication, including UAV relay, UAV jammer, AN, beamforming, and Reconfigurable Intelligent Surfaces (RIS).

\subsection{UAV-enabled Relaying and Jamming Schemes}
In V2X communication, two challenging scenarios arise. Firstly, when V2X communication occurs on different roads, the underlying links become NLoS, resulting in low received signal powers and reduced communication quality. Secondly, when V2X communication takes place on the same road over long distances, other vehicles on the road can act as obstacles and potential eavesdroppers, disrupting LoS links between legitimate transmitter-receiver pairs. In response to these challenges, an UAV can serve as aerial mobile relay to enhance communication quality, as illustrated in Fig. \ref{fig:scenario3}.
The optimal trajectory and beamforming matrix for the UAV must be carefully designed, taking into account the possibility of the UAV establishing LoS links with eavesdropping vehicles and the risk of information leakage caused by the side-lobes of the beams \cite{Zeng2016CM}. Additionally, the UAV can function as a friendly jammer to minimize the intercept probability of eavesdropping vehicles. However, in this scenario, the design of the UAV jammer's transmit power and trajectory also requires careful consideration to avoid increasing the outage probability of V2X communication \cite{Zhang2019TWC}.

\subsection{AN Generation in Orthogonal Channels}
Different cooperative jamming supported by UAV, when the number of antennas at legitimate transmitting vehicle is greater than that of eavesdropping vehicles, and CSI between legitimate transmitting vehicle and its receiving vehicle is known in advance, AN can be transmitted in the null space of legitimate channel for improving the likelihoods of achieving a positive secrecy capacity \cite{Liu2017TIFS}. That is $\mathrm{\mathbf{H_a Z}} = 0$ and $\mathrm{\mathbf{H_e Z}}\neq 0$, where $\mathrm{\mathbf{Z}}$ is the precoding matrix of the AN signal and known by legitimate vehicles. In this way, only the wiretap channel is degraded, and the legitimate channel does not be affected.

\subsection{Millimeter-Wave based Beamforming Design}
Apart from AN empowered by multiple antenna design, when vehicles or RSUs are equipped with a substantial number of antennas operating in the millimeter-wave frequency range, they can collaboratively degrade the channel quality of eavesdropping vehicles. This is achieved through the design of millimeter-wave beamforming features with narrow directivity, as illustrated in Fig. \ref{fig:scenario4}. By employing this strategy, the disparity between the quality of the legitimate channel and that of the eavesdropping channel is increased, thereby enhancing the secrecy performance.

\subsection{Joint design of RIS and UAV}
Integrating UAV communications with RIS represents an effective method to significantly enhance the secrecy performance of V2X communication, particularly in an UAV-enabled relaying system as depicted in Fig. \ref{fig:scenario3}. In situations where two RISs are deployed, it becomes possible to achieve a positive secrecy rate, even when the eavesdropping channel surpasses the channel between the relaying UAV and the receiving vehicle.
For example, as illustrated in Fig. \ref{fig:scenario1}, when a UAV equipped with an RIS is part of the system, it can serve as a friendly jammer to protect secret information by transmitting AN signals toward malicious vehicles. Simultaneously, the RIS installed on buildings can act as a relay, securely forwarding the confidential information to its intended destination. This coordinated use of RISs and the UAV enhances security and secrecy in V2X communication. Also, as presented in Fig. \ref{fig:scenario2}, the BS and its legitimate receiving vehicle face signal disconnections, reducing the effectiveness of PLS schemes. In such cases, a mobile RIS-based UAV can be an effective solution by providing LoS communication, helping to overcome signal blockages and maintain secure and reliable communication between the BS and its intended receiving vehicle.
In details, the RIS can use different phase shifts to produce constructive signal and destructive signal to a legitimate receiving vehicle and to an eavesdropping vehicle, respectively \cite{JavedTII2022, WangSPL2020, Ren2022TCOM}. However, a major challenge that hinders the widespread adoption of PLS supported by RIS and UAV is the difficulty of obtaining perfect CSI of eavesdropping vehicles. In this regard, more research on secure V2X communication is needed when combining RIS and UAVs, one of feasible methods is joint optimization of RIS-based beamforming and relaying/jamming UAV's trajectory.
Such joint optimization can potentially mitigate the impact of imperfect CSI and improve the overall performance of secure V2X communication systems.

While there has been significant progress in understanding how PLS can enhance the confidentiality and security of traditional wireless systems, it remains crucial to address the challenges that arise when PLS is adopted in V2X.
\begin{itemize}
  \item PLS heavily depends on the accurate knowledge of the underlying wireless channels. Furthermore, PLS schemes often make assumptions about the static positions of legitimate vehicles and eavesdropping vehicles. The mobility of vehicles and the time-varying nature of the wireless channel should be considered when implementing PLS, particularly in scenarios like V2X communication.
  \item The ideal assumption that the channel between legitimate transmitter-receiver vehicles is stronger than that between legitimate transmitting vehicles and malicious vehicles may not always hold true in practical scenarios due to vehicular mobility. This dynamic nature of vehicles can disrupt the feasibility of PLS strategies.
  \item Current PLS schemes have primarily concentrated on achieving security through the exploitation of communication interference. However, the intricate relationship between communication interference and sensing interference has not been given into consideration in the context of PLS for V2X communication. The impact of sensing interference among vehicles on the secrecy performance is an aspect that deserves more attention from researchers. Recognizing and addressing interference coupling can contribute significantly to enhancing the security of V2X communication systems.
\end{itemize}

\section{Open Issues and Future Research Directions}\label{sec:open issues and future work}
In this section, we discuss the open issues about PLS and their potential implications in V2X communications, including joint design of communication interference and sensing interference, superiority of RIS-enabled PLS, and joint design of vehicular trajectory prediction and UAV trajectory.

\subsection{Available CSI of PLS in V2X}
Effective relaying and jamming beamformings require accurate CSI of legitimate vehicles and eavesdropping vehicles. However, due to high mobility and vehicular flexibility, the CSI is time-varying and becomes imperfect. Consequently, relaying beamforming may be pointed to eavesdropping vehicles while jamming beamforming may be pointed to legitimate vehicles. In ISAC-enabled V2X systems, transmitting vehicles send signals that are used for communication with other devices and also for sensing their surroundings, under which signals reflected back to the transmitter contain information about the vehicle, like distance, speed, and direction. By using advanced signal processing techniques, ISAC can extract vehicular CSI from these reflected signals, encompassing critical parameters such as path loss, fading characteristics, and Doppler shifts.
By utilizing this extracted vehicular CSI, it becomes feasible to design relaying and jamming beamforming strategies that can be highly effective, even in the face of challenges arising from the time-varying and imperfect nature of CSI.

\subsection{Performance Restriction Relationship among Secure-ISAC}
\begin{figure}[!ht]
\centering
\includegraphics[width=2.6in]{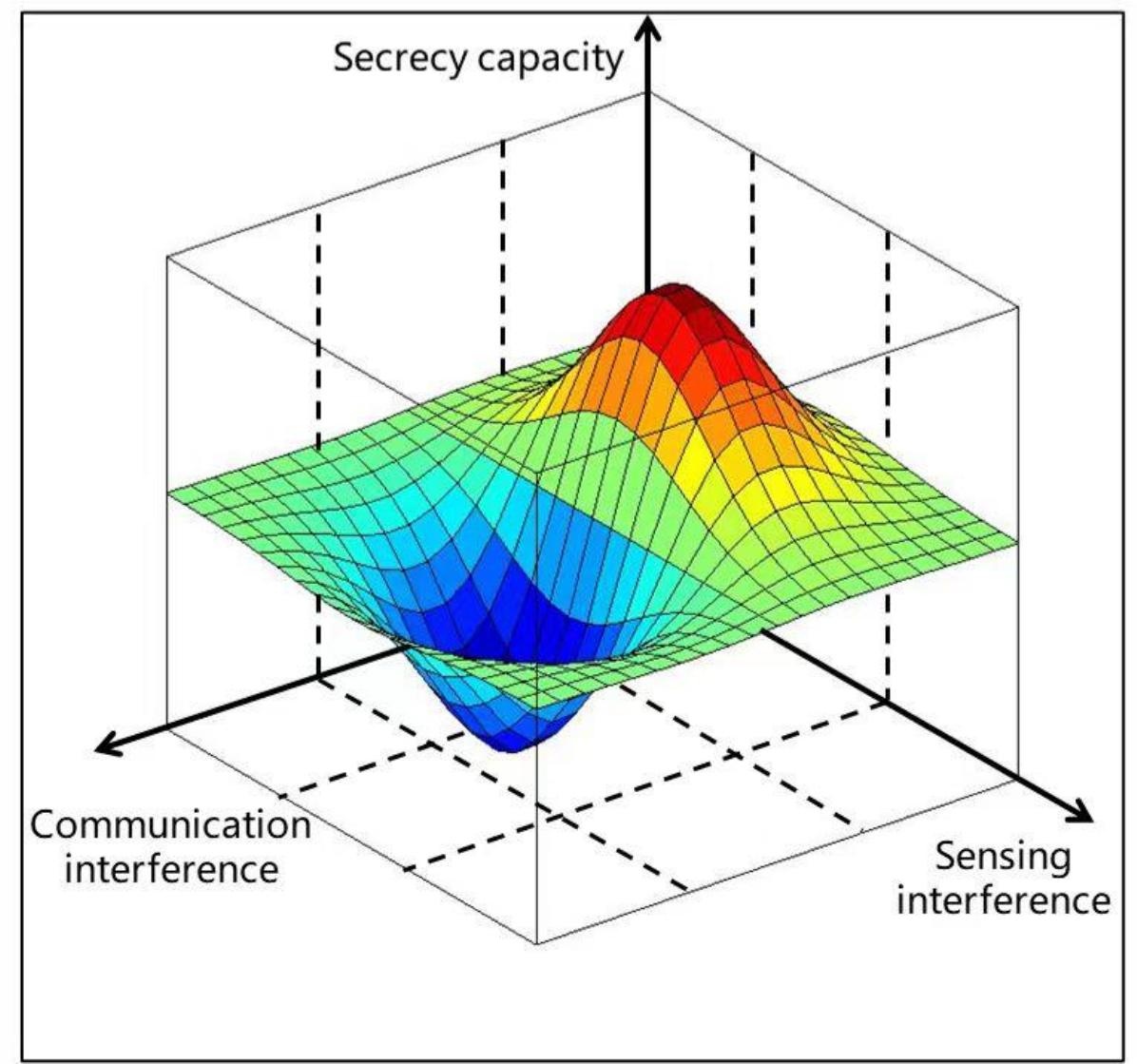}
\caption{\small The metric relationship among reliability, sensing, and security}
\label{fig:performance_ISACS}
\end{figure}

In the context of PLS systems within ISAC-based V2X, the use of sensing beamformings introduces a new dimension for PLS implementation. These sensing beamformings can serve as an innovative approach by replacing traditional interference techniques like AN and jamming.
Specifically, sensing beamformings can act as a form of ``green interference'', effectively suppressing wiretap channels without disrupting legitimate channels. What's particularly noteworthy is that this approach differs from traditional PLS design paradigms. Sensing beamforming not only enhances security but also improves reliability simultaneously. Therefore, it becomes highly effective to deeply integrate the design of sensing beamforming with communication beamforming. This joint design strategy helps effectively mitigate mutual interference and can be scheduled in a way that ensures the security of V2X communication is upheld while bolstering its reliability. Performance relationship among communication, sensing and security is explored in Fig. \ref{fig:performance_ISACS}.

Let $I_{\max}$ represent the maximum allowable interference at legitimate receiving vehicle. From Fig. \ref{fig:performance_ISACS}, it can be observed that given the reliability constraint, sensing interference can be utilized to maximize the secrecy capacity. In particular, as sensing interference continues to increase, secrecy capacity gradually decreases because it degrades the quality of legitimate channels more than that of eavesdropping channels. While the interference received at the legitimate receiving vehicle exceeds $I_{\max}$, the reliability constraint will be violated. Therefore, careful design of the sensing beamforming and communication beamforming needs to be crucial.

\subsection{Vehicle Trajectory Prediction for Securing V2X}
Many previous works on UAV trajectory design (or with the consideration of transmit power and block-length) for securing V2X communications assume that the locations of legitimate vehicles and eavesdropping vehicles are fixed, and optimize the UAV's horizontal trajectory. However, the mobility characteristics of vehicles has key effects on the PLS and should be designed carefully. The information of locations and CSI are two key factors for relaying and jamming beamforming design. Therefore,  joint design of vehicle trajectory prediction and UAV trajectory is an important research topic. Besides, joint design of vehicle trajectory prediction and sensing beamforming are also important to investigate. For example, legitimate transmitter can predict the vehicles¡¯ trajectory in adjacent and opposite lane, and send confidential message. In the meantime, exploiting multiple sensing beamforming supported by those vehicles in adjacent and opposite lane suppresses eavesdropping vehicles is also effective.

\subsection{Secure Cooperative Relaying for RIS-Enhanced V2X}
When V2X communication occurs on the same road with a long communication distance, other vehicles moving on the same road act as blockages and even eavesdropping vehicles. To improve the quality of legitimate channel effectively, the RIS mounted on top of the building can act as a friendly relay to forward the confidential message.
%Moreover, IRS-enhanced relay can be used to tune the phase shift of the signals smartly. As a result, the original and reflected signal can be added constructively to the vehicle.
Furthermore, RIS-enabled relaying beamforming for legitimate transmitter-receiver and sensing beamformings supported by other vehicles can be design jointly. That is, the beamformings of confidential message at RIS and sensing interference at other vehicles are pointed to the vehicle and malicious vehicles, respectively. In the future, there is a need to investigate the precision of beamforming and trade-off among reliability, sensing and security. However, several challenges still need to be investigated. For example, available CSI of legitimate vehicles and eavesdropping vehicles, large processing latency, and computational complexity.

\subsection{Non-negligible Sensing Interference in V2X}
\begin{figure}[!ht]
\centering
\includegraphics[width=3.2in]{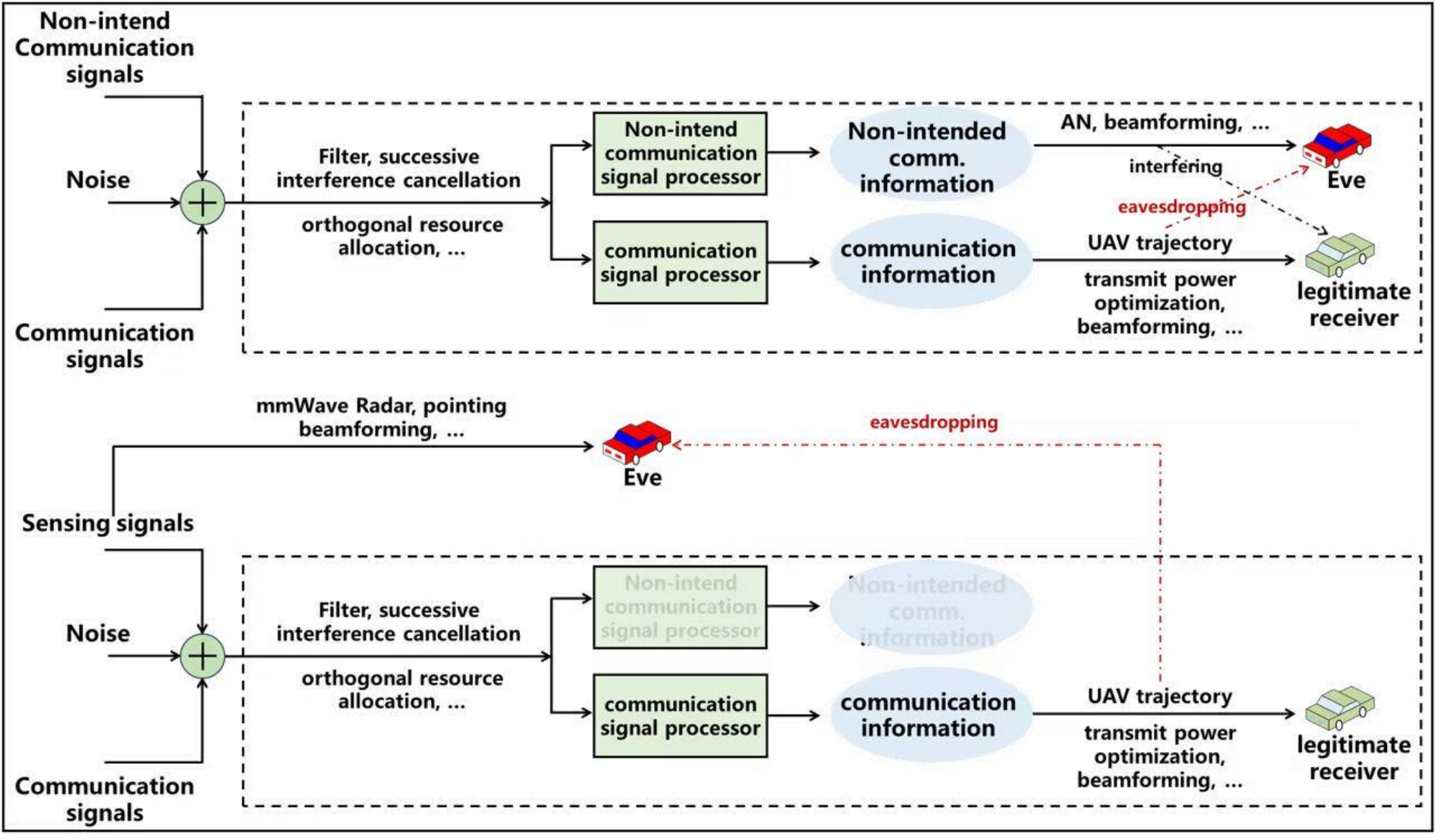}
\caption{\small PLS enhancement of V2X enabled by sensing function}
\label{fig:function_sensing_communication_security}
\end{figure}

The fundamental distinction between traditional PLS schemes and the potential of leveraging sensing beamforming in ISAC-based V2X communication is well-captured in Fig. \ref{fig:function_sensing_communication_security}. It's crucial to highlight that in ISAC-based V2X, communication and sensing signals coexist, and their performances are intertwined. Thus, the shift from traditional communication interference to sensing interference as a means to suppress eavesdropping channels presents an intriguing and innovative approach to enhance security. Focusing more attention on securing V2X communication by harnessing sensing interference is a promising avenue, and it opens up new avenues for research in this field.
\begin{itemize}
  \item the interplay between communication, sensing, and security in V2X communication is a complex and critical aspect that deserves thorough exploration. Shifting from a ``communication-enabled PLS'' approach to a more comprehensive ``joint communicating and sensing PLS'' paradigm could potentially revolutionize how we address security in V2X systems. This approach could offer greater insights into the unique challenges and opportunities presented by the integration of communication and sensing capabilities in intelligent transportation systems. Research in this direction could lead to innovative solutions and improved security for V2X communication.
  \item Establishing an analytical PLS framework that incorporates sensing beamforming for securing V2X communication is indeed a significant research challenge. This framework should account for various factors, including vehicular mobility, channel characteristics, and security requirements. Developing efficient algorithms to dynamically schedule sensing beamformings among legitimate vehicles while ensuring that they are directed towards potential malicious vehicles rather than legitimate intended vehicle is a complex task but critical for enhancing security.
\end{itemize}
However, no research works has been performed on this crucial topic from the perspective of sensing beamforming schedule. Therefore, it is preferred to explore novel methods in terms of sensing beamforming to protect against eavesdropping attacks.

In particular, for the scenario of Fig. \ref{fig:scenario4}, the sensing beamformings can be utilized to secure V2X communication based on the premise of their directionality. In details, vehicles in the forward direction of legitimate transmitting vehicle, locating in the adjacent opposite lane, employ forward sensing beamforming to suppress the quality of eavesdropping channels, with no impacts on other legitimate vehicles. Therefore, utilizing sensing beamforming from vehicles in adjacent opposite lanes, multiple vehicles collaboratively achieve eavesdropping channel suppression. Transmission reliability and security can be satisfied simultaneously.

\begin{figure}
\centering
\subfigure[\small Joint communication, sensing, and security for V2X]{
\centering
\includegraphics[width=3in]{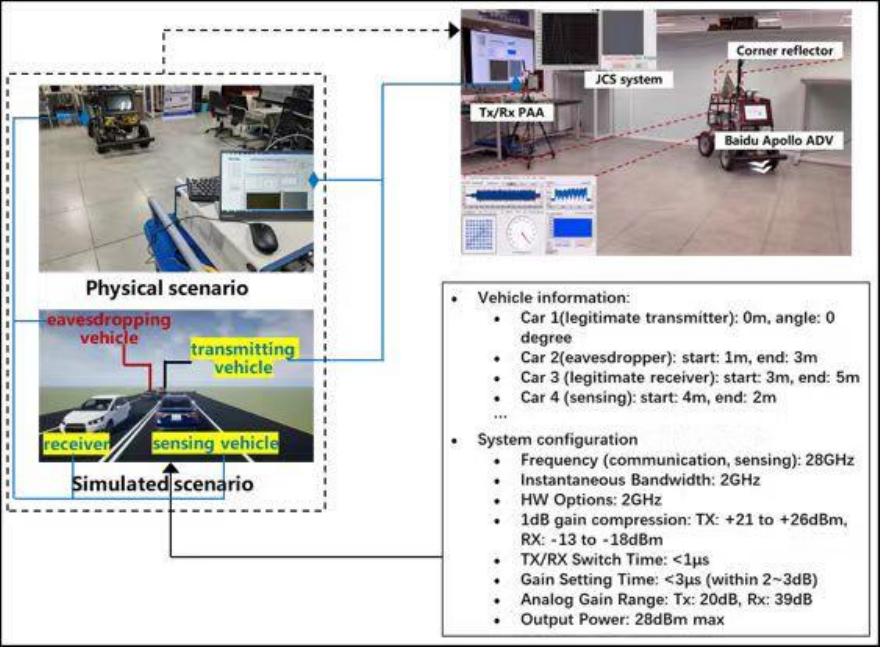}\label{fig:simulated_comm_sensing}}
\subfigure[\small Secrecy capacity empowered by sensing interference ]{
\centering
\includegraphics[width=3.1in]{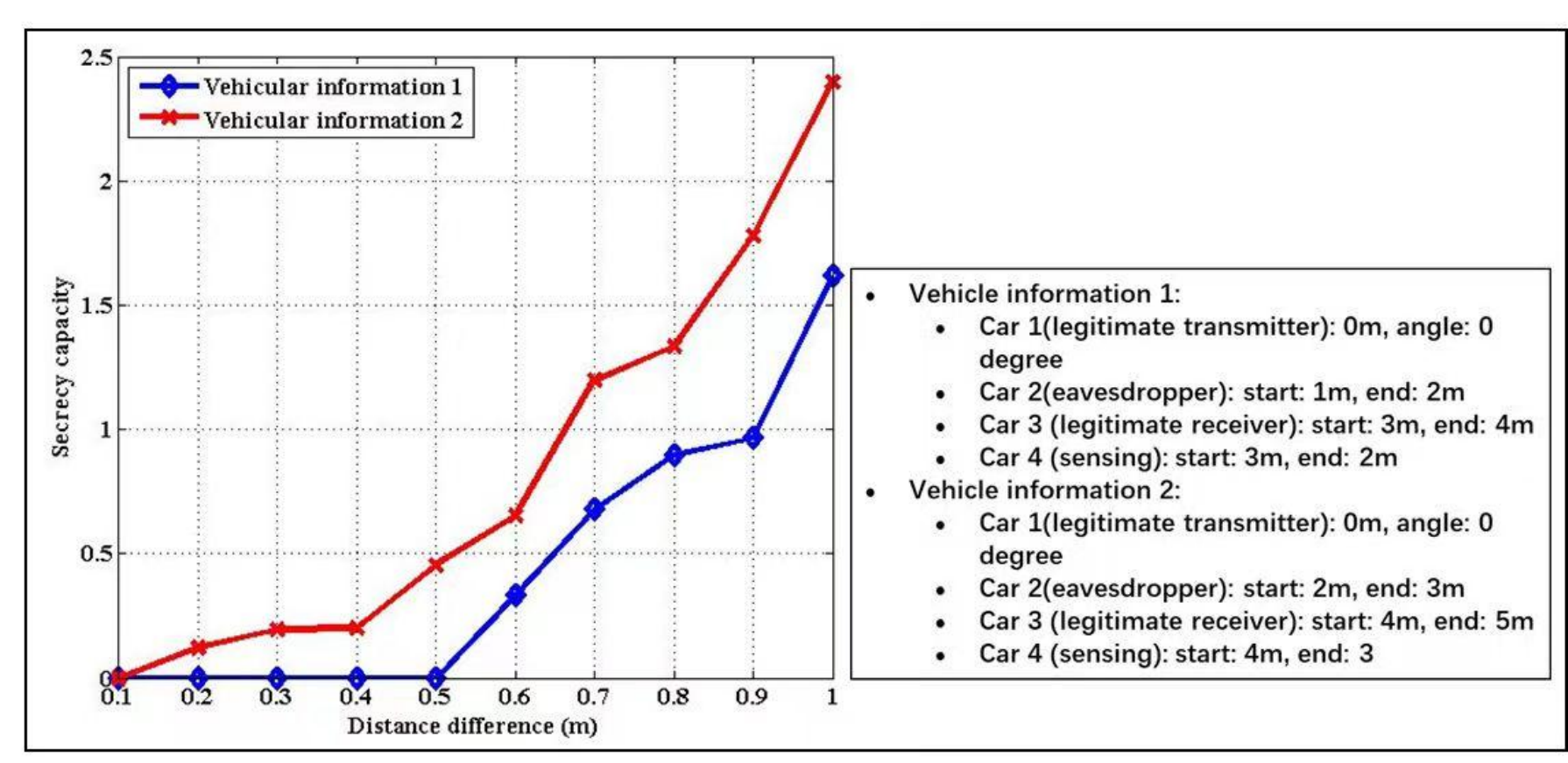}\label{fig:xingneng_systemparameters}}
\caption{\small SISAC hardware verification system and secrecy capacity test}
\end{figure}

\section{Conclusion}
In this article, we proposed a PLS enhancement framework by joint communication and sensing function in ISAC-enabled V2X. We outlined the challenges faced by current PLS techniques when applied to secure V2X communication. Furthermore, we discussed the role of sensing narrow-beamforming in the performance relationship of communication, sensing, and security, and pinpointed its enabling technologies. Numerical results verified the reliability, sensing success, and security of our proposed SISAC framework. We hope that our work will spur interests and open new directions for intelligent, efficient and secure V2X systems.

%\section*{Acknowledgements}
%This work is supported by the National Natural Science Foundation of China (61672321, 61771289, 61832012, and 61701269).

%%%%%%%%%%%%%%%%%%%%%%%%%%%%%%%%%%%%%%%%%%%%%%%%%%%%%%%%%%%%%
%%                  The Bibliography                       %%
%%                                                         %%
%%  Bmc_mathpys.bst  will be used to                       %%
%%  create a .BBL file for submission.                     %%
%%  After submission of the .TEX file,                     %%
%%  you will be prompted to submit your .BBL file.         %%
%%                                                         %%
%%                                                         %%
%%  Note that the displayed Bibliography will not          %%
%%  necessarily be rendered by Latex exactly as specified  %%
%%  in the online Instructions for Authors.                %%
%%                                                         %%
%%%%%%%%%%%%%%%%%%%%%%%%%%%%%%%%%%%%%%%%%%%%%%%%%%%%%%%%%%%%%

% if your bibliography is in bibtex format, use those commands:
\bibliographystyle{IEEEtran} % Style BST file (bmc-mathphys, vancouver, spbasic).
\bibliography{mybib}

\end{document}